\documentstyle[eqsecnum,preprint,psfig,aps]{revtex}
\tightenlines
\def\beq{\begin{equation}}
\def\eeq{\end{equation}}
\def\bea{\begin{eqnarray}}
\def\eea{\end{eqnarray}}
\def\nnu{\nonumber}
\def\eno#1{Eq.~(\ref{#1})}

\def\tst{\textstyle}
\def\by{\over}
\def\tofro{\leftrightarrow}

\def\Gam{\Gamma}
\def\Dta{\Delta}

\def\al{\alpha}

\def\gam{\gamma}

\def\tta{\theta}

\def\lam{\lambda}

\def\apx{\approx}
\def\quar{{\tst{1 \over 4}}}

\def\bH{{\bf H}}

\def\bS{{\bf S}}
\def\xhat{\bf{\hat x}}
\def\zhat{\bf{\hat z}}
\def\nhat{\bf{\hat n}}

\def\Fe8{Fe$_8$}

\def\Mn12{Mn$_{12}$}
\def\dadd{1.8mm}
\def\dadda{1.5mm}
\def\daddb{1mm}
\def\dsub{0mm}

\def\ham{{\cal H}}
\def\ket#1{|#1\rangle}
\def\mel#1#2#3{\langle#1|#2|#3\rangle}

\def\citn#1#2#3#4#5{#1, #2 {\bf#3}, #4 (#5)} 
\begin{document}

\draft


\title{
Diabolical Points in Molecular Magnets with a Four-Fold Easy Axis}

\author{Chang-Soo Park$^*$ and Anupam Garg$^{**}$}

\address{Department of Physics and Astronomy, Northwestern University,
Evanston, Illinois 60208}

\date{\today}
\maketitle

\begin{abstract}
We study the points of degeneracy (diabolical points) in magnetic
molecules such as Mn$_{12}$-acetate that have an easy axis of four-fold
symmetry. This is done for general magnetic field that need not be
oriented along a high-symmetry direction. We develop a perturbative
technique that gives the diabolical points as the roots of a small
number of polynomials in the transverse component of the magnetic field
and the fourth order basal plane anisotropy. In terms of these roots
we obtain approximate
analytic formulas that apply to any system with total spin
$S \le 10$. The analytic results are found to compare reasonably well
with exact numerical diagonalization for the case of \Mn12. In addition,
the perturbation theory shows that the diabolical points may be indexed
by the magnetic quantum numbers of the levels involved, even at large
transverse fields. Certain points of degeneracy are found to be mergers
(or near mergers) of two or three diabolical
points beacuse of the symmetry of the problem.
\end{abstract}

\pacs{75.10Dg, 03.65.Db, 03.65.Sq, 75.45.+j, 75.50.Xx} 

\section{INTRODUCTION}
\label{intro}

The magnetic molecules \Mn12 [short for Mn$_{12}$-acetate, or
[Mn$_{12}$(CH$_3$COO)$_{16}$(H$_2$O)$_4$O$_{12}$]
$\cdot$2CH$_3$COOH$\cdot$4H$_2$O]
and \Fe8 [short for Fe$_8$O$_2$(OH)$_{12}$(tacn)$_6$]$^{8+}$]
are among a few dozen that are currently being studied as extreme cases of
superparamagnets \cite{cds99}. Both \Mn12 and \Fe8 have spin 10, and both of
them display hysteresis
at the molecular level \cite{ns94,fs96,tlbd96,sop97}, as do some of the others.
In addition, however, \Fe8 shows an effect that has not yet been seen in any of
the other molecules: oscillation of the Landau-Zener-St\"uckelberg transition rate
between low lying levels as a function of the applied static magnetic field \cite{ws99}.
This oscillation is due to an oscillatory quenching of the underlying tunneling matrix
element connecting the levels in question, and is an unambiguous signature of
quantum tunneling. The occurrence and observability of tunneling in a spin of
such a large magnitude is of much interest in itself, and the tunneling frequency,
of the order of 100 Hz, is perhaps the lowest ever inferred or meaasured in any
physical system. Further, the quenching effect can be regarded as due to the interference
of different Feynman tunneling paths for the spin \cite{Loss1}, and this is how
it was first discovered \cite{Garg1}. While massive particle tunneling in two or
more spatial dimensions can also show such interference \cite{mw86}, the effect
arises more directly in the spin problem since the kinetic term in the action
has the mathematical structure of a Berry phase. This adds to the interest in
the problem. Reciprocally, the experimental observations have
motivated more careful investigations of spin-coherent-state
path integrals which are more delicate than their massive particle counterparts
\cite{spg00}.

The above intereference effect has also been sought in \Mn12, but has
not yet been seen. While the spin Hamiltonian for \Fe8 has biaxial symmetry,
\Mn12 is tetragonal, hence the systematics of the effect are different, and it
is interesting to calculate them. More
specifically, with an external magnetic field $\bH$,
\Mn12 is described by an anisotropy Hamiltonian \cite{Barra}
\beq
{\cal H} = -AS_z^2 - BS_z^4 + C(S_{+}^4 + S_{-}^4 ) -g\mu_b \bS\cdot\bH, \label{mn12}
\eeq
where $S=10$, $g \apx 2$, and 
$A \gg B \gg C > 0$. (The experimental values for $A$, $B$, and $C$ are
0.556, $ 1.1 \times 10^{-3}$, and $3 \times 10^{-5}$~K, respectively, so that
$\lam_1 = B/A = 1.98 \times 10^{-3}$, $\lam_2 = C/A =
5.4 \times 10^{-5}$, and $H_c = A/g\mu_B = 0.414$~T.)
The easy axis is now $z$ with four fold symmetry,
the hard axes are $\pm x$ and $\pm y$, and the medium axes are the
lines $y = \pm x$ in the $xy$-plane. For values of $|\bH|$ that are not too
large, the spin coherent state \cite{fnscs} expectation value
$\mel{\nhat}{\ham}{\nhat}$, which may be regarded as a classical energy $E(\tta,\phi)$,  
may have two or more local minima, between which the spin can then tunnel. The
issue is to calculate the tunnel splitting $\Dta$, and especially the field
values where this splitting is quenched, i.e., $\Dta = 0$. Since there may be
other molecules with this symmetry, it is desirable to do this analytically,
for all values of (or at least a wide range of) $\lam_1$ and $\lam_2$.

For the discussion that follows, it is useful to review some basic facts
pertaining to degeneracy in quantum mechanics in the absence of symmetry.
A point of degeneracy, or
equivalently, a point where the splitting between two levels vanishes, 
is {\it diabolical \/} in the terminology of Berry and Wilkinson \cite{diabolo2},
or a {\it conical intersection\/} in older terminology \cite{diabolo1}. As a rule,
eigenvalues of a finite Hamiltonian are all simple, and for a general Hamiltonian,
represented by a complex Hermitean matrix, we must be able to adjust at least
three parameters in order to produce a degeneracy. A simple approximate
argument is as follows \cite{va}.
Let two states $m$ and $m'$ be approximately degenerate,
and let the secular matrix between them be written as
\beq
\left(
    \begin{array}{cc}
        E_m  & V_{mm'} \\
        V_{m'm}       & E_{m'} 
    \end{array}
\right),    \label{sec}
\eeq
with $V_{m'm} = V^*_{mm'}$. The states will be truly degenerate only if
the following two conditions are met:
\bea
B_{mm'} &\equiv& E_m - E_{m'} = 0, \label{nobias}\\
V_{mm'} &=& 0. \label{nomix}
\eea
It is convenient to refer to these as the no-bias and the no-mixing conditions,
respectively \cite{vf}. Since $V_{mm'}$ is in general
complex, we have three real conditions, requiring three or more variable
parameters for their satisfaction.

Precisely three parameters
are available to us in \eno{mn12} in the three components of $\bH$. If the
Hamiltonian matrix is real symmetric, the number of adjustable parameters required
is lowered to two. In the present problem, this situation is realized when $H_y= 0$,
and so, as in the \Fe8 case, we expect to find the degeneracies in the $H_x$-$H_z$
plane. Unless explicitly stated, we will henceforth take $H_y=0$, so that
$V_{mm'} = V_{m'm}$.

Ignoring an additive constant, the energy surface in the vicinity of the degeneracy
is given by
\beq
E = \pm (B_{mm'}^2 + V_{mm'}^2)^{1/2}
\eeq
which has the form of a double cone or a {\it diabolo\/} in $H_x$-$H_z$ space,
which explains the term diabolical for these points.

In a previous paper \cite{Park}, we have studied the \Mn12 problem for $\bH\|\xhat$.
In this case, $E(\tta,\phi)$ is reflection symmetric
in the equatorial plane, and for small enough $H$, has two degenerate minima,
one in each hemisphere with $S_z > 0$ and $S_z < 0$. The problem is analogous to the
tunneling of a massive particle in a symmetric double well. The approach used is
a discrete phase integral (DPI) or Wentzel-Kramers-Brillouin (WKB) method, and it
provides a good quantitative approximation for the tunnel splitting,
$\Dta$, as well as the diabolical fields where the splitting vanishes.

In this paper, we extend our studies to the case where the field has a nonzero
$z$ component. The problem is now like a massive particle in an asymmetric double well.
Our initial intent was to approach this case also using the DPI method, as it has been
appplied successfully to the \Fe8 problem. Indeed, the calculation reduces to no more
than the evaluation of a handful of action integrals, and based on prior experience
with \Fe8 and other tunneling Hamiltonians, we are confident that it will be
quantitatively accurate for \Mn12 too. Unlike the \Fe8 case, however,
where the integrals can be found analytically, this turns out to be not so for
\Mn12, and so, even though the calculation is not as atomistic as a brute force
diagonalization of the $21\times 21$ Hamiltonian matrix, it still does not yield
final answers in an analytic form. To our pleasant surprise, however, we have
found that a perturbative approach in the spirit of Ref.~\cite{gargdia}
not only yields quite good approximations
for the locations of the diabolical points, it also reveals a pattern in the
points which would be hard to discern from a numerical diagonalization
by itself. Most importantly, it provides a scheme for indexing these
points. It is the purpose of this paper to report this work.

The analysis is contained in Sec.~\ref{pert}. We first develop
the perturbation theory with $B = 0$ (Sec.~\ref{Beq0}). This yields the
diabolical points as the roots of polynomials in $H_x$ and $C$. For a given spin
$S$, we have $2S$ polynomials. One of the
unexpected bonuses is that a polynomial which applies to a given value of $S$
also applies to any other value of $S$. We find all these polynomials for
$S \le 10$. In subsection \ref{Bneq0}, we incorporate the effects of the
$B$ term approximately, and compare our analytic results with those from
explicit numerical diagonalization of the Hamiltonian with the parameters for
\Mn12 (see Table~\ref{alldps}). Our results are accurate to about 10\%
for \Mn12, and we believe that they will also be useful for other systems
with four-fold anisotropy. This is especially true for the low lying energy
levels. In fact, for the higher pairs of levels, the diabolical points can
be significantly moved or even eliminated altogether by still
higher ansitropy terms in the Hamiltonian. 
In subsection \ref{merger} we discuss some qualitative
aspects of the degeneracies on the $H_z$ axis, and show that some of them
behave as the merger of two or three diabolical points. A short summary
(Sec.~\ref{summ}) concludes the paper.

\section{Perturbative Calculation}
\label{pert}

Let us first consider the problem when $H_x=0$, i.e., $\bH\|\zhat$.
The $S_{\pm}^4$ terms couple states with $m$ differing by 4, and thus
divide the Hamiltonian into four disjoint subspaces. 
Levels belonging
to different subspaces can cross at fields which can be approximately
obtained by neglecting the $S_{\pm}^4$ terms. The crossing
conditions are $E_m = E_{m'}$, where $E_m = -Am^2 - Bm^4
- g\mu_B H_z m$. Improved formulas can be obtained by finding
corrections to $E_m$ perturbatively in $C$. These intersections
are easy to understand because a symmetry of the Hamiltonian (invariance
under a rotation by 90$^{\circ}$ about $\zhat$) is easily
recognized.

If $H_x$ and $H_z$ are both non-zero, there is no obvious symmetry.
If $H_x$ and $C$ are both small, however, we may continue to label
the states by the $m$ quantum numbers, and calculate the energies
perturbatively in these two parameters. In terms of the secular matrix
(\ref{sec}), the energies $E_m$, $E_{m'}$, and the bias $B_{mm'}$ are determined
by the terms in $S_z$ in \eno{mn12}, and the mixing $V_{mm'}$ by the
terms involving $C$ and $H_x$. The energies are trivial to find, so the
problem is to find $V_{mm'}$.

It is convenient to divide the Hamiltonian by $A$ and to work with
scaled quantities
\beq
\lam_1 = B/A, \quad \lam_2 = C/A,
       \quad h_{x,z} = H_{x,z} /S H_c, \label{redvar}
\eeq
with
\beq
H_c \equiv A/g\mu_B. \label{Hcdef}
\eeq 

\subsection{Simplified model: $B = 0$.}
\label{Beq0}

To keep the problem tractable, let us set $B=0$ at this stage. Then,
to zeroth order in both $C$ and $H_x$, $E_m = -Am^2 -g\mu_BH_z m$.  
Hence, levels $m$ and $m'$ are degenerate when
\begin{equation}
h_z = -\frac{m+m'}{S}. \label{hzcond}
\end{equation}
It remains to find the off-diagonal element
$V_{mm'}$. As we shall see, the choice $B = 0$ simplifies the
calculation greatly, for $E_m$ is then quadratic in $m$, and energy level
differences are linear in $m$, and given by a fixed set of numbers
whenever \eno{hzcond} holds.

To illustrate the calculation of $V_{mm'}$, we consider the case $S=5$.
Suppose $h_z = 1/5$, so
that $m = -5$ and $m' = 4$ are degenerate (see Fig.~\ref{levels}).
We will find $V_{4,-5}$ to leading non-zero order in $h_x$
and $C$ as a double series in these variables. It is clear that a
transition from $m$ to $m'$ can be made in three ways:

(1) act with $h_xS_+$ in 9th order.

(2) act with $h_xS_+$ in 5th order and $CS_+^4$in 1st order.

(3) act with $h_xS_+$ in 1st order and $CS_+^4$ in 2nd order.\\
We denote the corresponding contributions to $V$ by $V^{(1)}$,
$V^{(2)}$, and $V^{(3)}$. Each of these involves a product of
matrix elements and a product of energy denominators. For
$V^{(1)}$, the former is
\begin{equation}
\frac{1}{2^9}\langle5,4\mid S_+^9 \mid5,-5\rangle(-h_xS)^9 =
709\sqrt{10}(-h_xS)^9 \equiv W(-h_x S)^9. \label{matrix}
\end{equation}
In fact the number $W$ will be common (up to some power of 2) to
all three $V^{(i)}$. The energy denominators can be read off
Fig.~\ref{levels}, and for $V^{(1)}$, these are
\begin{equation}
(-1)^8 (8\times14\times18\times20)^2 \equiv (-1)^8K^2.
\label{denomi}
\end{equation}
The factor $(-1)^8$ appears here because all intermediate states
are higher in energy than $E_{-5}$ and $E_4$, and we have
introduced the number $K$ because many of the energy denominator
products for $V^{(2)}$ and $V^{(3)}$ contain the same factors.
Putting together Eqs.~(\ref{matrix}) and (\ref{denomi}),
\begin{equation}
V^{(1)} = -\frac{W}{K^2}(h_xS)^9.
\end{equation}

For $V^{(2)}$, the transition can occur in several ways, six to be
precise, corresponding to where the $S_+^4$ term acts. Two of the
ways to make the transition are $-5 \rightarrow -1 \rightarrow 0
\rightarrow 1 \rightarrow 2 \rightarrow 3 \rightarrow 4$, and $-5
\rightarrow -4 \rightarrow 0 \rightarrow 1 \rightarrow 2
\rightarrow 3 \rightarrow 4$. The product of matrix elements in
each case is
\begin{equation}
\left(-\frac{h_xS}{2}\right)^5\lam_2 <4\mid S_+^9 \mid -5> =
-2^4W\lam_2(h_xS)^5.
\end{equation}
The energy denominators, however, depend on the transition path,
and are listed in Table \ref{path}. Adding together all the
contributions, we get
\begin{eqnarray}
V^{(2)} &=& 2^4W\frac{2}{K}\left(\frac{1}{8} + \frac{1}{20} +
\frac{20}{112}\right)(h_xS)^4\lam_2\nonumber\\ &=& \frac{2^4
\times 99}{140}\frac{W}{K}(h_xS)^5\lam_2.
\end{eqnarray}

Lastly, for $V^{(3)}$, there are three transition paths: (i) $-5
\rightarrow -1 \rightarrow 3 \rightarrow 4$, (ii) $-5 \rightarrow
-1 \rightarrow 0 \rightarrow 4$, (iii) $-5 \rightarrow -4
\rightarrow 0 \rightarrow 4$. The transition element product for
all three is
\begin{equation}
-2^8 W(h_x S)\lam_2^2,
\end{equation}
and the energy denominator product is $20 \times 8$, $20 \times
20$, and $8 \times 20$, respectively. Thus,
\begin{equation}
V^{(3)} = -\frac{2^8 \times 3}{200}W(h_xS)\lam_2^2.
\end{equation}

Adding together $V^{(1)}$, $V^{(2)}$, and $V^{(3)}$, we obtain the
net $V_{4,-5}$ (restoring the level quantum numbers). The
condition for a diabolic point is that this quantity should
vanish. In addition to $h_x = 0$, this happens when
\begin{equation}
\xi^2 - \frac{99K}{140} \xi + \frac{3K^2}{200}
= 0, \label{quadra}
\end{equation}
where
\begin{equation}
\xi = {1 \over \lam_2} \left( \frac{h_xS}{2} \right)^4.
\end{equation}
Solving Eq.~(\ref{quadra}), we obtain
\begin{equation}
h_x = \frac{2}{5} \left\{ \frac{99}{280} \pm \left[
\left(\frac{99}{280}\right)^2 -
\frac{3}{200}\right]^{\frac12}\right\}^{\frac{1}{4}}(K\lam_2)^{\frac{1}{4}}.
\end{equation}
With the scaled value $\lam_2 = 2.16\times 10^{-4}$ for $S=5$, this
yields $h_x = 0.2643$, and 0.6252. Direct numerical
diagonalization yields $h_x = 0.2669$, and 0.638.

Readers will undoubtedly have noted that apart from an overall
factor of $h_x$ to some power, our perturbation method yields the
off-diagonal element as a homogeneous polynomial in $h_x^4$ and
$\lam_2$. It is not difficult to see that this will be
generally true, and also not difficult to justify. Let us first
take the point that we have only included transition paths that go
through the higher energy levels. Consider, e.g. the path for
$V_{4,-5}$ in the above calculation that goes from $-5$ to $+1$
via six successive $h_xS_+$ terms, then to $+5$ via a $CS_+^4$
term, and finally to $+4$ via an $h_xS_-$ term. This term is of
order $h_x^7\lam_2$, and should be regarded as a higher order
correction to $V^{(2)}$. Secondly, it is positive and of the same
sign as $V^{(2)}$, because it involves six negative and one
positive energy denominator. This feature is also generally valid,
and is important in light of the next point, which is that the
sign of the terms in the polynomial for the off-diagonal element
alternate when organized as a series in $\lam_2$. Thus, for
$V_{4,-5}$, $V^{(1)}$ is negative, $V^{(2)}$ is positive, and
$V^{(3)}$ is negative. This is a consequence of the fact that
replacing four $-h_xS_+$ terms by a single $CS_+^4$ term $(a)$
leaves the sign of the matrix element product unchanged, but $(b)$
replaces four negative energy denominators by a single negative
one.

Let us call the polynomial that remains after we have cancelled
off as many overall factors of $h_xS$ from $V_{mm'}$ as possible
the {\it underlying polynomial}. It is clear that this polynomial
is of degree
\begin{equation}
n_{mm'} = \left[ \left| \frac{m - m'}{4} \right| \right],
\end{equation}
in $h_x^4$, where $[x]$ denotes the integer part of $x$, i.e., the
largest integer less than or equal to $x$. The alternation of
signs of successive powers of $h_x^4$ is a necessary (but not
sufficient) condition for all $n_{mm'}$ roots to be positive
\cite{note4}. This means that, not including the points on
the $h_x$ or $h_z$ axes, it is possible for states labelled by $m$ and
$m'$ to intersect in a diabolical point {\it up to} $n_{mm'}$
times in the first quadrant of the $h_x$-$h_z$ plane. This appears
to us to be a topological property of the Mn$_{12}$ Hamiltonian,
that is not altered by presence of higher order terms, as long as the
symmetry is not changed. Of course,
the number of diabolical points may be fewer, but we do not
believe it can be greater, because if $h_x$ is sufficiently large,
the term $H_xS_x$ dominates the energy in the equatorial plane,
and the possibility of interfering trajectories is lost. We do not
have a proof of these statements, which must be regarded as
conjectures, but the similarity to Fe$_8$, and all the empirical
evidence we have gathered suggests that they are indeed true.

For $S=5$, we have found all the diabolical points using this
perturbation approach, and also 
numerically. In all cases, the perturbative answers are nearly
exact. The results are shown in Fig.~\ref{diabolic}.

At this point we wish to note a remarkable feature of this
approximation, which may have been noticed by some readers. This
is that the diabolical values of $h_x$ depend on $m$ and $m'$
only through the combination $\Dta m = m' - m$. This means that
in Fig.~\ref{diabolic}, the theoretical points corresponding to
the same value of $\Dta m = m' - m$ are vertically aligned. See, for
example, the points corresponding to $m = -4$, $m'=4$, and $m=-5$,
$m'=3$, or to $m=-5$, $m'=2$, and $m=-4$,$m'=3$. The reason is not
hard to find. When we set $B=0$, the energy $E_m$ is a quadratic
function of $m$, and when levels $m$ and $m' > m$ are degenerate,
\beq
E_{m+k} - E_m = k(\Dta m -k). \label{Emkm}
\eeq
Thus, {\it the entire pattern of energy levels
above the levels $m$ and $m'$ depends only on $\Dta m$\/} (see
Fig.~\ref{levels}),
and since only these levels enter into our perturbation theory, the
energy denominators are identical. The matrix elements are of course
different, but since our transition paths involve no closed loops, they
amount to a net factor of $\mel{m'}{S_+^{\Dta m}}{m}$ in each term,
which therefore drops out of the underlying polynomial. In short, this
polynomial depends only on $\Dta m$.

Furthermore, we see from \eno{Emkm} that value of $S$ also does not
enter in the energy denominators. This means that the polynomials
found for $S=5$ are applicable to the $S=10$ problem for transitions
with $\Dta m \le 10$, and makes it worthwhile to find all the remaining
polynomials for $S=10$. The task is easily automated on a computer since
the problem is essentially to enumerate the transition paths
for a given order in $h_x$ and $\lam_2$, and to add up the reciprocals
of the corresponding energy denominators. It can be further simplified
by noting that all denominators appear in the term proportional to
$h_x^{\Dta m}$, so that relative to this term, the coefficient for
any other transition path appears as an energy numerator, consisting
of all the missing denominators. 
The polynomials are given in Table~\ref{polytab} , along with the
roots for $h_x$ for general $S$ and $\lam_2$, as well as for the values
applicable to \Mn12.

One last general point worth noting is that for a diabolical point
labelled by the pair $(m,m')$ with $m'>m$, the energy levels which are
degenerate are numbers $2S + (m - m') + 1$ and $2S + (m - m') + 2$,
where the ground state is given the number 1.

\subsection{Inclusion of $S_z^4$ term}
\label{Bneq0}

When we try and compare the results of the previous subsection with those
obtained numerically from \eno{mn12} with $B\ne 0$, we find that the
systematics of the diabolical points are fully captured in that the analytic
results provide a complete indexing scheme, but the quantitative
disagreement with the \Mn12 parameters is as bad as $30\%$ in some cases.
We therefore seek some way to incorporate the $B\ne 0$ effects.

It is easy to let $B \ne 0$ in the no-bias condition. Equation
(\ref{hzcond}) is modified to
\beq
h_z(m,m') = -{1 \over S} (m + m')
          \left[   1  + \lam_1 (m^2 + m'^2) \right]. \label{hzmm}
\eeq
The no-mixing condition is clearly harder to evaluate. A simple minded
approach is to shift the energy levels so as to retain the same relative
spacings as when $B=0$, but allow the overall range to be modified. 

With this in mind, let us first consider the energies when $\lam_1=0$. 
When levels $m$ and $m'$ are degenerate, the level at the top of the
barrier is given by the quantum number $k = (m+m')/2$, whenever this
is an integer, or by the nearest integers if it is a half-integer. In
the spirit of our approximations, keeping track of this distinction
would be an overrefinement, so we will use the formula $(m + m')/2$ in
both cases. The energy range is thus given by
\beq
\Dta E^{(0)} = E^{(0)}_k - E^{(0)}_m = \quar (m-m')^2, \label{DE0}
\eeq
where the (0) superscript indicates that $B=0$. With $B \ne 0$, we get
\bea
\Dta E^{(1)} &=& \Dta E^{(0)}
             -\lam_1 \left[ \left({m+m'} \by 2 \right)^4 - m^4 \right]
             + \lam_1 (m + m')(m^2 + m'^2)
                    \left[ {m+m' \over 2} - m \right] \nnu \\
            &=& \gam_{mm'} \Dta E^{(0)}, \label{DE1}
\eea
where
\beq
\gam_{mm'} = 1 + {\lam_1 \over 4}(7m^2 + 10 mm' + 7 m'^2).
                   \label{gmm}
\eeq

If we assume that the entire spectrum gets modified from its quadratic
form by a uniform stretching factor $\gam_{mm'}$, then the only change in
our perturbation theory is that all energy denominators get multiplied by
this factor. In the underlying polynomial, $h_x$ and $\lam_2$ get
replaced by $h_x/\gam_{mm'}$ and $\lam_2/\gam_{mm'}$, and hence the
no-mixing condition becomes
\beq
h_x(m,m') = \gam_{mm'}^{3/4} r_{\al}(\Dta m),
    \quad \al = 1, 2, \ldots, n_{mm'}, \label{hxmm}
\eeq
where $\{r_{\al}\}$ are the original $h_x$ values obtained from the roots of the
underlying polynomial $P_{\Dta m}$.

The formulas (\ref{hzmm}) and (\ref{hxmm}) are compared with exact numerical
results in Table~\ref{alldps}. The errors are now typically about 10\%, and
can be of either sign.

It is useful to briefly discuss our numerical procedure. For points lying
on the $H_z$ or $H_x$ axis, the splitting is a function of one variable,
and its zeros can be found by simple scanning. For the off-axis zeros, this
is harder, and we resort to the
Herzberg and Longuet-Higgins sign change theorem \cite{diabolo1,mvb84},
which applied to the present problem states that, upon adiabatic traversal
of a closed contour in the $H_x$-$H_z$ plane enclosing a single point of
degeneracy of two states, the wavefunction of either of these two states
returns to itself except for a change in sign.  Conversely,
there is no change in sign if the contour does not enclose a degeneracy. Hence,
to find a diabolical point, we first find a sign-reversing rectangular
contour by hit and trial. By bisecting this rectangle in the $x$ and $z$
directions alternately, and using the sign-change test at each bisection,
we can corral the degeneracy ever more tightly, to the degree desired.
We have found this procedure to be generally superior to a direct
minimization of the energy difference for the reason that the diabolo in the
vicinity of a degeneracy is highly asymmetrical in the $x$ and $z$ directions.
Consider for example, the 4th and 5th energy levels from the bottom when
$h_z \simeq 0.12$--$0.13$, corresponding approximately to the $m$ quantum
numbers $-9$ and $8$. Since these states are separated by a high barrier,
the mixing element between them is best understood as arising from tunneling,
and will therefore be proportional to an exponentially small Gamow factor
$e^{-\Gam}$, where $\Gam$ is the appropriate tunneling action. 
Thus the energy surface consists of a deep and narrow valley running nearly
parallel to the $h_x$ axis, with a valley floor that goes to zero linearly
at occasional points, and may rise and fall in between. Because of this shape,
and because the surface is not analytic in the vicinity of the points being
sought, standard methods for finding the minima of a function are often not well
suited.

The above argument also enables to understand an observation made by Berry
and Wilkinson \cite{diabolo1} and Berry and Mondragon \cite{bm86} in the
study of two very different model problems, namely, that the
energy cone near a diabolical point has very high eccentricity in terms
of the physically natural parameters describing the system. In other words,
the cross section of the energy surface is a very long and narrow ellipse
in one direction. Let us see how this happens in the present problem.
To save writing let us write just $x$ and $z$ for the deviations of $h_x$
and $h_z$ from a diabolical point at $(h_{x0},h_{z0})$. In the vicinity of
this point, we have
\bea
B_{mm'} &\simeq& a z , \label{biasexp} \\
V_{mm'} &\simeq& b e^{-\Gam} x , \label{mixexp}
\eea
where $a$ and $b$ are constants of order unity. Thus the energy surface is
\beq
E \simeq (a^2  z^2 + b^2 e^{-2\Gam} x^2)^{1/2}. \label{Esurf}
\eeq
The cross section is an ellipse with major axis parallel to $x$ and eccentricity
$\sim e^{\Gam} \gg 1$. This scenario is expected to be quite general. The
no-bias condition defines a line in parameter space. The gradient of the bias
normal to this line is generally expected to be of order unity
in the natural physical variables. The mixing element also varies on the
same order unity scale in the parameter space, but because it arises from
tunneling, its absolute scale is very small. The result is an energy surface of
high eccentricity of the type just described.

\subsection{Merged diabolical points [22]}
\label{merger}

Our discussion of diabolical points needs some elaboration for certain
degeneracies lying on the $h_z$ axis. For the pair $(m,m') = (-10,9)$, e.g.,
(more generally any pair with $\Dta m = 4n + 3$), we peeled off a factor
of $h_x^3$ from the mixing element. Writing $x = h_x$, and $z = h_z$ as in
\eno{Esurf}, and $z_0 = h_{z0}$ for the point of degeneracy, the
bias and mixing are given by
\bea
B_{mm'} &\apx& z - z_0, \label{bias2}\\
V_{mm'} &\apx& x^3 + O(x^7), \label{mix2}
\eea
ignoring multiplicative constants. Correspondingly, the energy surface
is $[(z-z_0)^2 + x^6]^{1/2}$, whose cross section is no longer an ellipse.
There is also no reason for the simple sign-change result to hold a priori. 

These conclusions are based on perturbation
theory, however. More generally, we can only argue on grounds of symmetry that,
under $x \to -x$, $B_{mm'} \to B_{mm'}$, and $V_{mm'} \to \pm V_{mm'}$.
Instead of Eqs.~(\ref{bias2}) and (\ref{mix2}), we should therefore expect
the general expansion to take the form
\bea
B_{mm'} &\apx& z - z_0 + a x^2 
             + O\left( (z-z_0)^2, x^2(z-z_0)^2, x^4 \right), \label{bias3}\\
V_{mm'} &\apx& x^3 - b x + c x (z-z_0)
             + O\left( x^3(z-z_0), x^5 \right), \label{mix3}
\eea
where $a$, $b$, and $c$ are constants, all of which we expect to be very small
on account of the quantitative accuracy of the perturbative approach.
Ignoring the higher order terms, the bias vanishes on the parabola
$z = z_0 - a x^2$. On this parabola, the mixing is given by
\beq
(1-ac)x^3 - b x, \label{mix4}
\eeq
which vanishes at $x=0$, and $x = \pm [b/(1-ac)]^{1/2} \equiv \pm x_1$,
assuming that $b/(1-ac) > 0$.
Thus instead of a single degeneracy at $(0,z_0)$, we have three closely
spaced degeneracies at
\beq
(0,z_0), \quad (\pm x_1, z_0 - ax_1^2). \label{tripdiab}
\eeq
The energy surface in the immediate vicinity of each one of these points
is now diabolical in the ordinary sense. A small circuit of each of these
points separately will therefore lead to a sign reversal, as will a larger
circuit enclosing all three of them. If we ignore the splitting, we may
regard the original degeneracy on the $h_z$ axis as a {\it triply merged}
diabolical point.

It is useful to think of the
coefficients $a$, $b$, and $c$ as depending on parameters in the Hamiltonian
other than the components of $\bH$, e.g., $\lam_1$ and $\lam_2$. It may be
that as these parameters are varied, the quantity $b/(1-ac)$ becomes
negative, so that the roots $\pm x_1$ cease to be real. We can think of
the two off-axis diabolical points as having annihilated each other, leaving
behind only one true diabolical point on the axis. Unless one is very close
to this point, however, the energy surface may still resemble that of a triply
merged point. 

For the parameters appropriate to \Mn12, we find that the points are
located at \cite{fnabc}
\beq
(h_x, h_z) = (0.0, 0.135836224),
           \quad (\pm 0.01855, 0.135832551). \label{tdloc}
\eeq
These numbers are obtained
by using the same sign-reversal theorem as previously described. Because
the energy difference
depends so sensitively on $h_z$, however, we have confirmed them in
another way. For any given value of $h_x$, we first
find the minimum of the relevant energy gap $\Dta$ with respect to $h_z$.
In essence, we find the value of the gap at the bottom of the parabolic
trench where the bias vanishes. A plot of this gap versus $h_x$ should be
given by the absolute value of the expression (\ref{mix4}). As shown in
Fig.~\ref{mergedps}, this is indeed so, and the split off point is again
found to be at $h_x = 0.01855$.

In exactly the same way, we may also consider {\it doubly merged} points,
corresponding to the tunneling of states with $\Dta m = 4n +2$, e.g.,
$(m,m') = (-10,8)$, when $h_x \apx 0$.  We can continue to expand the bias
as in \eno{bias3}, but the leading term in $V_{mm'}$ is now proportional to
$x^2$, so that instead of \eno{mix3}, we have
\beq
V_{mm'} \apx x^2 + O\left( (z-z_0), x^2(z-z_0), x^4 \right). \label{mix5}
\eeq
It is then obvious that the double zero of $V_{mm'}$ at $x=0$ can not be
split. This conclusion can also be reached in another way. Symmetry would
require that, if they split, the points be located at $(\pm x_0, z_0)$,
with $x_0 \ne 0$. The selection rule argument given at the very beginning
of Sec.~\ref{intro} shows, however, that this is impossible, as there must
be a crossing of any two levels $m$ and $m'$ with $\Dta m \ne 4n$ as $H_z$
is varied with $H_x = 0$. In Fig.~\ref{mergedps}, we also show the
$(m,m') = (-10,8)$ gap at the bottom of the no-bias trench. It is apparent
that now the diabolical points remain unsplit at $h_x = 0$, and the curve
is extremely well fit by a parabola, as required by \eno{mix5}.

Finally, the points on the $h_z$ axis, corresponding to $\Dta m = 4n + 1$
are singly diabolical to begin with, so the issue of splitting does not arise.

\section{Summary}
\label{summ}
We have studied the diabolical points of a spin Hamiltonian describing
molecules such as \Mn12 that have an easy axis of four-fold
symmetry, for arbitrarily directed magnetic fields. A perturbation theory
in the parameters $H_x$ and $C$ is found to give a very good qualitative
and even quantitative understanding. Our central results are the formulas
(\ref{hzmm}) and (\ref{hxmm}), which along with the results
in Table~\ref{polytab}
give the full set of diabolical points for any molecule with $S \le 10$.

\acknowledgments
We are indebted to Wolfgang Wernsdorfer for very useful correspondence.
AG's research is supported by the NSF via grant number DMR-9616749.

\newpage

\begin{figure}
\caption{(a) Energy level diagram for $S=5$, with $B=0$, and
$h_z = 1/5$. (b) Part of diagram for $S=10$ with $h_z = 0.3$, showing
that the pattern above any two degenerate levels depends only on
$\Dta m$, irrespective of $S$.}
\label{levels}
\end{figure}

\begin{figure}
\caption{Diabolical points for $S=5$. For $\lam_2$ we have used
the scaled value $2.16 \times 10^{-4}$. Each point is labeled by
the Zeeman quantum numbers $(m,m')$ except those on the $h_z$ axis.
Note that points with the same value of $m+m'$ are horizontally
aligned, while those with the same $\Dta m = m' - m$ are vertically
aligned. For the points on the $h_z$ axis, any pair $(m,m')$ is
allowed, consistent with the given value of $m+m'$ and the rule
$\Dta m \ne 4n$.}
\label{diabolic}
\end{figure}

\begin{figure}
\caption{Merged and nearly merged diabolical points. The plot shows
the tunnel splitting $\Dta(m,m')$, showing how the triply
merged [$(m,m') = (-10,9)$] point is split, but the doubly
merged [$(m,m') = (-10,8)$] point is not. The splitting is calculated
along the bottom of the parabolic trench in the $h_x$-$h_z$ plane.
In other words, for each value of $h_x$, what is plotted is the
minimum of the splitting with respect to $h_z$. For the
$-10 \tofro 9$ triple merger,
also plotted is the curve $\Dta = |\al h_x (1 - h^2_x/h^2_{x1})|$, with
$\al = 1.51 \times 10^{-12}$, and $h_{x1} = 0.01855$, but this curve
cannot be distinguished from the points on the scale of this figure,
on account of the size and density of the symbols. (They can be
distinguished on a large computer screen.) Similarly, the $-10 \tofro 8$
splitting is very accurately fit to a parabola.}
\label{mergedps}
\end{figure}

\begin{table}[ht]
\caption{Transition paths and energy denominators for perturbative
calculation of $V^{(2)}$.} \vspace{0.2cm}
\begin{center}
\begin{tabular}{l|l}
Transition path & Energy denominator product\\ \hline $-5
\rightarrow -1 \rightarrow 0 \rightarrow 1 \rightarrow 2
\rightarrow 3 \rightarrow 4$ &$(-1)^5 20K$ \\ $-5 \rightarrow -4
\rightarrow 0 \rightarrow 1 \rightarrow 2 \rightarrow 3
\rightarrow 4$ &$(-1)^5 8K$\\ $-5 \rightarrow -4 \rightarrow -3
\rightarrow 1 \rightarrow 2 \rightarrow 3 \rightarrow 4$ &$(-1)^5
(8\times 14/20)K$\\ $-5 \rightarrow -4 \rightarrow -3 \rightarrow
-2 \rightarrow 2 \rightarrow 3 \rightarrow 4$ &$(-1)^5 (8\times
14/20)K$\\ $-5 \rightarrow -4 \rightarrow -3 \rightarrow -2
\rightarrow -1 \rightarrow 3 \rightarrow 4$ &$(-1)^5 8K$\\ $-5
\rightarrow -4 \rightarrow -3 \rightarrow -2 \rightarrow -1
\rightarrow 0 \rightarrow 4$ & $(-1)^5 20K$
\end{tabular}
\end{center}
\label{path}
\end{table}

\begin{table}
\caption{Underlying polynomials and $h_x$ values for $S=10$. In
columns 1 and 2, $\Delta m = m' - m$ $(m' > m)$, and $\xi =
\lam_2^{-1}(Sh_x/2)^4$. The quantity $g$ is a convenient multiplier 
that enables reduction of the coefficients.
Column 4 gives the fourth root of the roots of $P_{\Delta m}(\xi)$,
i.e., the quantity $Sh_x/\lam_2^{1/4}$. In
column 5, we give the $h_x$ values for $\lam_2 = 5.4
\times 10^{-5}$ and $S=10$.} \vspace{0.2cm}
\begin{center}
\begin{tabular}{c|l|c|l|l}
$\Dta m$ & $P_{\Dta m}(\xi)$ & $g$ & $Sh_x/\lam_2^{1/4}$ & $h_x$ \\[\daddb]
\hline
4 & $\xi - 36$ & --- & 4.8990 & 0.04200 \\[\daddb]
\hline
5 & $\xi - 288 $ & --- & 8.2391 & 0.07063 \\[\daddb]
\hline
6 & $\xi - 1296$ & --- & 12.000 & 0.1028 \\[\daddb]
\hline
7 & $\xi - 4320$ & --- & 16.214 & 0.1390 \\[\daddb]
\hline
8 & $\xi^2 - 66g \xi + 49g^2 $ & 180 & 6.8195,
20.821 & 0.05846, 0.1785 \\[\daddb]
\hline
9 & $\xi^2 - 99 g\xi + 294 g^2$ & 288 & 10.901,
25.785 & 0.09345,  0.2210 \\[\daddb]
\hline
10 &$\xi^2 - 429 g\xi + 9604 g^2$ & 144 & 15.286,
31.086 & 0.1310, 0.2665 \\[\daddb]
\hline
11 &$\xi^2 -
429 g\xi + 13854 g^2$ & 288 & 20.065,
36.703 & 0.1720, 0.3146\\[\daddb]
\hline
12 &
 $\xi^3 - 1287 g\xi^2 + 162171 g^2\xi -266805 g^3 $ &180 &
8.3243, 25.185, 42.620 & 0.07136, 0.2159, 0.3654\\[\daddb]
\hline
13 &$\xi^3
 - 286 g\xi^2 + 9767 g^2\xi - 11858 g^3 $ &1440 &
 13.054, 30.620, 48.822 & 0.1119, 0.2625, 0.4185 \\[\daddb]
\hline
14 & $\xi^3 - 4862 g\xi^2 + 3297473 g^2\xi - 140612164 g^3$ & 144
 &18.013, 36.357, 55.296 &  0.1544, 0.3116, 0.4740 \\[\daddb]
\hline
15 &$\xi^3 - 3978 g\xi^2 + 2501703 g^2\xi - 144597726 g^3$ & 288
 &23.327, 42.383, 62.033 & 0.2000, 0.3633, 0.5318 \\[\daddb]
\hline
16 &$\xi^4 - 50388 g\xi^3 + 444908598 g^2\xi^2$ &36
&9.600, 28.943,      &  0.08230, 0.2481, \\[\dsub]
&\ $-474703246836 g^3\xi + 6904413140625 g^4$ &
&48.684, 69.022      &  0.4173, 0.5917\\[\daddb]
\hline
17 &$\xi^4 -
3876 g\xi^3 + 2869532 g^2\xi^2$ & 720 &
14.907, 34.841,       & 0.1278, 0.2987, \\[\dsub]
&\ $ -315894768 g^3\xi + 1301534080 g^4$ &  &
55.251, 76.254        & 0.4736, 0.6537 \\[\daddb]
\hline
18 &$\xi^4 - 5814 g\xi^3 + 6946761 g^2\xi^2$
&720 & 20.389, 41.013,    &   0.1748, 0.3516, \\[\dsub]
&\ $-1452656484 g^3\xi + 20248151616 g^4$ & &
62.073, 83.721            &   0.5321, 0.7177   \\[\daddb]
\hline
19 &$\xi^4 -
 21318 g\xi^3 + 99462359 g^2\xi^2 $ &288 &
 26.201, 47.448,          &   0.2246, 0.4067, \\[\dsub]
 &\  $-92627838402 g^3\xi + 8455413407896 g^4 $
 & & 69.141, 91.417       &   0.5927, 0.7837 \\[\daddb]
\hline
20 & $ \xi^5 - 245157 g\xi^4 + 13893314634 g^2\xi^3
 $ &36 &10.727, 32.285,   &   0.09196, 0.2768, \\[\dsub]
 &\  $ -175140030572298 g^3\xi^2$ & & 54.138, 76.448,
 &0.4641, 0.6553, \\[\dsub]
 &\ $+ 285990169496161221 g^4\xi$
 & & 99.335 &0.8515 \\[\dsub]
 &\ $ -648297466934390625 g^5 $ & & &
\end{tabular}
\end{center}
\label{polytab}
\end{table}

\newpage
\begin{table}
\caption{Analytic and numerical results for diabolical point
locations for the \Mn12 Hamiltonian. For each pair $(m,m')$, the
upper line lists the modified perturbation theory answers for $(h_x,h_z)$
as given by Eqs.~(\ref{hxmm}) and (\ref{hzmm}), and the lower
line lists the numerical results. We have not separated the merged
diabolical points located on the $h_z$ axis (Sec.~\ref{merger}).}
\begin{center}
\begin{tabular}{c | l l l l l}
$(m,m')$ & \multicolumn{5}{c}{$(h_x,h_z)$} \\
\hline
$(-10, 10)$ & (0.1053, 0.0) & (0.3169, 0.0)  & (0.5314, 0.0)
                               & (0.7504, 0.0)  & (0.9751, 0.0)  \\[\dsub]
          & (0.1161, 0.0) & (0.3440, 0.0)  & (0.5681, 0.0)
                               & (0.7888, 0.0)  & (1.0051, 0.0)  \\[\dadd]
$(-9,   9)$ & (0.0, 0.0)    & (0.1954, 0.0)  & (0.3931, 0.0)
                               & (0.5949, 0.0)  & (0.8024, 0.0)  \\[\dsub]
          & (0.0, 0.0)    & (0.2097, 0.0)  & (0.4161, 0.0)
                               & (0.6206, 0.0)  & (0.8234, 0.0)  \\[\dadd]
$(-8,   8)$ & (0.0900, 0.0) & (0.2713, 0.0)  & (0.4564, 0.0)
                               & (0.6471, 0.0)  &   \\[\dsub]
          & (0.0953, 0.0) & (0.2836, 0.0)  & (0.4711, 0.0)
                               & (0.6587, 0.0)  &   \\[\dadd]
$(-7,   7)$ & (0.0, 0.0)    & (0.1655, 0.0) & (0.3341, 0.0)  & (0.5081, 0.0)  & \\[\dsub]
          & (0.0, 0.0)    & (0.1703, 0.0) & (0.3400, 0.0)  & (0.5110, 0.0)  &  \\[\dadd]
$(-6,   6)$ & (0.0751, 0.0) & (0.2273, 0.0)  & (0.3847, 0.0)  & & \\[\dsub]
          & (0.0759, 0.0) & (0.2273, 0.0)  & (0.3803, 0.0)  & & \\[\dadd]
$(-5,   5)$ & (0.1359, 0.0)  & (0.2763, 0.0)  & & & \\[\dsub]
          & (0.1328, 0.0)  & (0.2668, 0.0)  & & & \\[\dadd]
$(-4,   4)$ & (0.0598, 0.0)  & (0.1827, 0.0)  & & & \\[\dsub]
          & (0.0569, 0.0)  & (0.1708, 0.0)  & & & \\[\dadd]
$(-3,   3)$ & (0.1042, 0.0)  & & & & \\[\dsub]
          & (0.0931, 0.0)  & & & & \\[\dadd]
$(-2,   2)$ & (0.0422, 0.0)  & & & & \\[\dsub]
          & (0.0353, 0.0)  & & & & \\[\dadd]
$(-1,   1)$ & (0.0, 0.0)  & & & & \\[\dsub]
          & (0.0, 0.0)  & & & & \\[\dadd]
\hline
$(-10,  9)$ & (0.0, 0.1358)    & (0.2546, 0.1358)  & (0.4610, 0.1358)
                        & (0.6717, 0.1358)  & (0.8882, 0.1358)  \\[\dsub]
           & (0.0, 0.1358)    & (0.2747, 0.1350)  & (0.4904, 0.1332)
                        & (0.7036, 0.1304)  & (0.9139, 0.1263)  \\[\dadd]
$(-9,   8)$ & (0.0, 0.1287)     & (0.1415, 0.1287)  & (0.3308, 0.1287)
                        & (0.5246, 0.1287)  & (0.7240, 0.1287)  \\[\dsub]
          & (0.0, 0.1287)     & (0.1507, 0.1284)  & (0.3480, 0.1273)
                        & (0.5445, 0.1252)  & (0.7403, 0.1221)  \\[\dadd]
$(-8,   7)$ & (0.0, 0.1224)     & (0.2169, 0.1224)  & (0.3941, 0.1224)
                        & (0.5767, 0.1224)  & \\[\dsub]
          & (0.0, 0.1224)     & (0.2250, 0.1217)  & (0.4040, 0.1204)
                        & (0.5838, 0.1182)  & \\[\dadd]
$(-7,   6)$ & (0.0, 0.1168)     & (0.1191, 0.1168)  & (0.2794, 0.1168)
                        & (0.4454, 0.1168)  & \\[\dsub]
          & (0.0, 0.1168)     & (0.1215, 0.1166)  & (0.2819, 0.1159)
                        & (0.4444, 0.1146)  & \\[\dadd]
$(-6,   5)$ & (0.0, 0.1121)     & (0.1801, 0.1121)  & (0.3294, 0.1121)  & & \\[\dsub]
          & (0.0, 0.1121)     & (0.1781, 0.1118)  & (0.3220, 0.1112)  & & \\[\dadd]
$(-5,   4)$ & (0.0, 0.1081)     & (0.0964, 0.1081)  & (0.2281, 0.1081)  & & \\[\dsub]
          & (0.0, 0.1081)     & (0.0932, 0.1081)  & (0.2170, 0.1081)  & & \\[\dadd]
$(-4,   3)$ & (0.0, 1050)       & (0.1418, 0.1050)  & & & \\[\dsub]
          & (0.0, 1049)       & (0.1299, 0.1052)  & & & \\[\dadd]
$(-3,   2)$ & (0.0, 0.1026)     & (0.0714, 0.1026)  & & & \\[\dsub]
          & (0.0, 0.1026)     & (0.0625, 0.1028)  & & & \\[\dadd]
$(-2,   1)$ & (0.0, 0.1010)  & & & & \\[\dsub]
          & (0.0, 0.1009)  & & & & \\[\dadd]
$(-1,   0)$ & (0.0, 0.1002)  & & & & \\[\dsub]
          & (0.0, 0.1001)  & & & & \\[\dadd]
\hline
$(-10,  8)$ & (0.0, 0.2649)     & (0.1969, 0.2649)  & (0.3961, 0.2649)
                        & (0.5995, 0.2649)  & (0.8085, 0.2649)  \\[\dsub]
          & (0.0, 0.2649)     & (0.2108, 0.2640)  & (0.4186, 0.2610)
                        & (0.6249, 0.2561)  & (0.8298, 0.2491)  \\[\dadd]
$(-9,   7)$ & (0.0907, 0.2515)  & (0.2735, 0.2515)  & (0.4600, 0.2515)
                        & (0.6522, 0.2515)  & \\[\dsub]
          & (0.0958, 0.2513)  & (0.2854, 0.2495)  & (0.4744, 0.2461)
                        & (0.6638, 0.2409)  & \\[\dadd]
$(-8,   6)$ & (0.0, 0.2396)     & (0.1669, 0.2396)  & (0.3368, 0.2396)
                        & (0.5122, 0.2396)  & \\[\dsub]
          & (0.0, 0.2396)     & (0.1715, 0.2389)  & (0.3425, 0.2368)
                        & (0.5150, 0.2334)  & \\[\dadd]
$(-7,   5)$ & (0.0758, 0.2293)  & (0.2292, 0.2293)  & (0.3879, 0.2293)  & & \\[\dsub]
          & (0.0765, 0.2291)  & (0.2290, 0.2282)  & (0.3833, 0.2264)  & & \\[\dadd]
$(-6,   4)$ & (0.0, 0.2206)     & (0.1370, 0.2206)  & (0.2787, 0.2206)  & & \\[\dsub]
          & (0.0, 0.2206)     & (0.1339, 0.2203)  & (0.2690, 0.2198)  & & \\[\dadd]
$(-5,   3)$ & (0.0604, 0.2135)  & (0.1843, 0.2135)  & & & \\[\dsub]
          & (0.0574, 0.2134)  & (0.1724, 0.2138)  & & & \\[\dadd]
$(-4,   2)$ & (0.0, 0.2079)     & (0.1052, 0.2079)  & & & \\[\dsub]
          & (0.0, 0.2079)     & (0.0940, 0.2085)  & & & \\[\dadd]
$(-3,   1)$ & (0.0426, 0.2040)  & & & & \\[\dsub]
          & (0.0357, 0.2042)  & & & & \\[\dadd]
$(-2,   0)$ & (0.0, 0.2016)  & & & & \\[\dsub]
          & (0.0, 0.2014)  & & & & \\[\dadd]
\hline
$(-10,  7)$ & (0.0, 0.3885)     & (0.1437, 0.3885)  & (0.3359, 0.3885)
                        & (0.5327, 0.3885)  & (0.7353, 0.3885)  \\[\dsub]
          & (0.0, 0.3885)     & (0.1525, 0.3877)  & (0.3523, 0.3844)
                        & (0.5520, 0.3782)  & (0.7518, 0.3962)  \\[\dadd]
$(-9,   6)$ & (0.0, 0.3695)     & (0.2203, 0.3695)  & (0.4003, 0.3695)
                        & (0.5859, 0.3695)  & \\[\dsub]
          & (0.0, 0.3695)     & (0.2279, 0.3677)  & (0.4097, 0.3637)
                        & (0.5929, 0.3574)  & \\[\dadd]
$(-8,   5)$ & (0.0, 0.3529)     & (0.1210, 0.3529)  & (0.2839, 0.3529)
                        & (0.4527, 0.3529)  & \\[\dsub]
          & (0.0, 0.3528)     & (0.1232, 0.3523)  & (0.2861, 0.3502)
                        & (0.4514, 0.3464)  & \\[\dadd]
$(-7,   4)$ & (0.0, 0.3386)     & (0.1831, 0.3386)  & (0.3349, 0.3386)  & & \\[\dsub]
          & (0.0, 0.3385)     & (0.1809, 0.3377)  & (0.3273, 0.3361)  & & \\[\dadd]
$(-6,   3)$ & (0.0, 0.3267)     & (0.0981, 0.3267)  & (0.2320, 0.3267)  & & \\[\dsub]
          & (0.0, 0.3267)     & (0.0948, 0.3266)  & (0.2209, 0.3266)  & & \\[\dadd]
$(-5,   2)$ & (0.0, 0.3172)     & (0.1443, 0.3172)  & & & \\[\dsub]
          & (0.0, 0.3171)     & (0.1324, 0.3180)  & & & \\[\dadd]
$(-4,   1)$ & (0.0, 0.3101)     & (0.0727, 0.3101)  & & & \\[\dsub]
          & (0.0, 0.3101)     & (0.0637, 0.3108)  & & & \\[\dadd]
$(-3,   0)$ & (0.0, 0.3053)  & & & & \\[\dsub]
          & (0.0, 0.3051)  & & & & \\[\dadd]
$(-2,  -1)$ & (0.0, 0.3030)  & & & & \\[\dsub]
          & (0.0, 0.3028)  & & & & \\[\dadd]
\hline
$(-10,  6)$ & (0.0928, 0.5077)  & (0.2799, 0.5077)  & (0.4708, 0.5077)
                        & (0.6674, 0.5077)  & \\[\dsub]
          & (0.0975, 0.5073)  & (0.2909, 0.5040)  & (0.4843, 0.4975)
                        & (0.6791, 0.4876)  & \\[\dadd]
$(-9,   5)$ & (0.0, 0.4840)     & (0.1709, 0.4840)  & (0.3449, 0.4840)
                        & (0.5245, 0.4840)  & \\[\dsub]
          & (0.0, 0.4839)     & (0.1750, 0.4826)  & (0.3499, 0.4787)
                        & (0.5270, 0.4722)  & \\[\dadd]
$(-8,   4)$ & (0.0776, 0.4634)  & (0.2349, 0.4634)  & (0.3974, 0.4634)  & & \\[\dsub]
          & (0.0781, 0.4631)  & (0.2342, 0.4613)  & (0.3926, 0.4579)  & & \\[\dadd]
$(-7,   3)$ & (0.0, 0.4459)     & (0.1405, 0.4459)  & (0.2856, 0.4459)  & & \\[\dsub]
          & (0.0, 0.4459)     & (0.1372, 0.4454)  & (0.2759, 0.4445)  & & \\[\dadd]
$(-6,   2)$ & (0.0619, 0.4317)  & (0.1890, 0.4317)  & & & \\[\dsub]
          & (0.0589, 0.4316)  & (0.1771, 0.4323)  & & & \\[\dadd]
$(-5,   1)$ & (0.0, 0.4206)     & (0.1079, 0.4206)  & & & \\[\dsub]
          & (0.0, 0.4205)     & (0.0968, 0.4217)  & & & \\[\dadd]
$(-4,   0)$ & (0.0437, 0.4127)  & & & & \\[\dsub]
          & (0.0368, 0.4131)  & & & & \\[\dadd]
$(-3,  -1)$ & (0.0, 0.4079)  & & & & \\[\dsub]
          & (0.0, 0.4077)  & & & & \\[\dadd]
\hline
$(-10,  5)$ & (0.0, 0.6238)     & (0.2272, 0.6237)  & (0.4128, 0.6237)
                        & (0.6042, 0.6237)  & \\[\dsub]
          & (0.0, 0.6237)     & (0.2338, 0.6209)  & (0.4212, 0.6147)
                        & (0.6110, 0.6049)  & \\[\dadd]
$(-9,   4)$ & (0.0, 0.5960)     & (0.1249, 0.5960)  & (0.2930, 0.5960)
                        & (0.4672, 0.5960)  & \\[\dsub]
          & (0.0, 0.5960)     & (0.1266, 0.5952)  & (0.2944, 0.5919)
                        & (0.4656, 0.5861)  & \\[\dadd]
$(-8,   3)$ & (0.0, 0.5723)     & (0.1890, 0.5723)  & (0.3458, 0.5723)  & & \\[\dsub]
          & (0.0, 0.5722)     & (0.1865, 0.5709)  & (0.3380, 0.5685)  & & \\[\dadd]
$(-7,   2)$ & (0.0, 5525)       & (0.1014, 0.5525)  & (0.2397, 0.5525)  & & \\[\dsub]
          & (0.0, 5524)       & (0.0979, 0.5523)  & (0.2285, 0.5523)  & & \\[\dadd]
$(-6,   1)$ & (0.0, 0.5366)     & (0.1491, 0.5366)  & & & \\[\dsub]
          & (0.0, 0.5365)     & (0.1373, 0.5379)  & & & \\[\dadd]
$(-5,   0)$ & (0.0, 0.5248)     & (0.0752, 0.5248)  & & & \\[\dsub]
          & (0.0, 0.5247)     & (0.0663, 0.5258)  & & & \\[\dadd]
$(-4,  -1)$ & (0.0, 0.5168)  & & & & \\[\dsub]
          & (0.0, 0.5165)  & & & & \\[\dadd]
$(-3,  -2)$ & (0.0, 0.5129)  & & & & \\[\dsub]
          & (0.0, 0.5126)  & & & & \\[\dadd]
\hline
$(-10,  4)$ & (0.0, 0.7378)     & (0.1775, 0.7378)  & (0.3582, 0.7378)
                        & (0.5448, 0.7378)  & \\[\dsub]
          & (0.0, 0.7377)     & (0.1808, 0.7359)  & (0.3622, 0.7306)
                        & (0.5470, 0.7217)  & \\[\dadd]
$(-9,   3)$ & (0.0807, 0.7069)  & (0.2441, 0.7069)  & (0.4132, 0.7069)  & & \\[\dsub]
          & (0.0809, 0.7065)  & (0.2429, 0.7041)  & (0.4080, 0.6995)  & & \\[\dadd]
$(-8,   2)$ & (0.0, 0.6808)     & (0.1461, 0.6808)  & (0.2972, 0.6808)  & & \\[\dsub]
          & (0.0, 0.6807)     & (0.1426, 0.6802)  & (0.2873, 0.6790)  & & \\[\dadd]
$(-7,   1)$ & (0.0644, 0.6594)  & (0.1967, 0.6594)  & & & \\[\dsub]
          & (0.0614, 0.6594)  & (0.1849, 0.6604)  & & & \\[\dadd]
$(-6,   0)$ & (0.0, 0.6428)     & (0.1123, 0.6428)  & & & \\[\dsub]
          & (0.0, 0.6427)     & (0.1015, 0.6443)  & & & \\[\dadd]
$(-5,  -1)$ & (0.0456, 0.6309)  & & & & \\[\dsub]
          & (0.0387, 0.6316)  & & & & \\[\dadd]
$(-4,  -2)$ & (0.0, 0.6238)  & & & & \\[\dsub]
          & (0.0, 0.6235)  & & & & \\[\dadd]
\hline
$(-10,  3)$ & (0.0, 0.8511)     & (0.1306, 0.8511)  & (0.3064, 0.8511)
                        & (0.4886, 0.8511)  & \\[\dsub]
          & (0.0, 0.8510)     & (0.1317, 0.8501)  & (0.3070, 0.8459)
                        & (0.4867, 0.8385)  & \\[\dadd]
$(-9,   2)$ & (0.0, 0.8178)     & (0.1979, 0.8178)  & (0.3620, 0.8178)  & & \\[\dsub]
          & (0.0, 0.8177)     & (0.1949, 0.8162)  & (0.3451, 0.8131)  & & \\[\dadd]
$(-8,   1)$ & (0.0, 0.7901)     & (0.1062, 0.7901)  & (0.2511, 0.7901)  & & \\[\dsub]
          & (0.0, 0.7900)     & (0.1026, 0.7899)  & (0.2400, 0.7901)  & & \\[\dadd]
$(-7,   0)$ & (0.0, 0.7679)     & (0.1563, 0.7679)  & & & \\[\dsub]
          & (0.0, 0.7678)     & (0.1448, 0.7697)  & & & \\[\dadd]
$(-6,  -1)$ & (0.0, 0.7513)     & (0.0788, 0.7513)  & & & \\[\dsub]
          & (0.0, 0.7512)     & (0.0701, 0.7528)  & & & \\[\dadd]
$(-5,  -2)$ & (0.0, 0.7402)  & & & & \\[\dsub]
          & (0.0, 0.7399)  & & & & \\[\dadd]
$(-4,  -3)$ & (0.0, 0.7347)  & & & & \\[\dsub]
          & (0.0, 0.7344)  & & & & \\[\dadd]
\hline
$(-10,  2)$ & (0.0849, 0.9647)  & (0.2570, 0.9647)  & (0.4348, 0.9647)  & & \\[\dsub]
          & (0.0848, 0.9643)  & (0.2551, 0.9614)  & (0.4296, 0.9559)  & & \\[\dadd]
$(-9,   1)$ & (0.0, 0.9299)     & (0.1539, 0.9299)  & (0.3131, 0.9299)  & & \\[\dsub]
          & (0.0, 0.9298)     & (0.1501, 0.9292)  & (0.3032, 0.9278)  & & \\[\dadd]
$(-8,   0)$ & (0.0679, 0.9014)  & (0.2074, 0.9014)  & & & \\[\dsub]
          & (0.0648, 0.9014)  & (0.1959, 0.9027)  & & & \\[\dadd]
$(-7,  -1)$ & (0.0, 0.8792)     & (0.1185, 0.8792)  & & & \\[\dsub]
          & (0.0, 0.8791)     & (0.1080, 0.8812)  & & & \\[\dadd]
$(-6,  -2)$ & (0.0481, 0.8634)  & & & & \\[\dsub]
          & (0.0414, 0.8643)  & & & & \\[\dadda]
$(-5,  -3)$ & (0.0, 0.8539)  & & & & \\[\dsub]
          & (0.0, 0.8536)  & & & & \\[\dadda]
\hline
$(-10,  1)$ & (0.0, 1.0800)     & (0.2095, 1.0800)  & (0.3832, 1.0800)  & & \\[\dsub]
          & (0.0, 1.0799)     & (0.2061, 1.0782)  & (0.3753, 1.0747)  & & \\[\dadd]
$(-9,   0)$ & (0.0, 1.0443)     & (0.1125, 1.0443)  & (0.2661, 1.0443)  & & \\[\dsub]
          & (0.0, 1.0443)     & (0.1089, 1.0442)  & (0.2553, 1.0444)  & & \\[\dadd]
$(-8,  -1)$ & (0.0, 1.0158)     & (0.1658, 1.0158)  & & & \\[\dsub]
          & (0.0, 1.0158)     & (0.1547, 1.0179)  & & & \\[\dadd]
$(-7,  -2)$ & (0.0, 0.9944)     & (0.0836, 0.9944)  & & & \\[\dsub]
          & (0.0, 0.9944)     & (0.0753, 0.9962)  & & & \\[\dadd]
$(-6,  -3)$ & (0.0, 0.9802)  & & & & \\[\dsub]
          & (0.0, 0.9799)  & & & & \\[\dadd]
$(-5,  -4)$ & (0.0, 0.9731)  & & & & \\[\dsub]
          & (0.0, 0.9729)  & & & & \\[\dadd]
\hline
$(-10,  0)$ & (0.0, 1.1980)     & (0.1638, 1.1980)  & (0.3331, 1.1980)  & & \\[\dsub]
          & (0.0, 1.1979)     & (0.1599, 1.1973)  & (0.3237, 1.1957)  & & \\[\dadd]
$(-9,  -1)$ & (0.0723, 1.1624)  & (0.2209, 1.1624)  & & & \\[\dsub]
          & (0.0693, 1.1624)  & (0.2099, 1.1638)  & & & \\[\dadd]
$(-8,  -2)$ & (0.0, 1.1346)     & (0.1263, 1.1346)  & & & \\[\dsub]
          & (0.0, 1.1346)     & (0.1164, 1.1369)  & & & \\[\dadd]
$(-7,  -3)$ & (0.0513, 1.1148)  & & & & \\[\dsub]
          & (0.0449, 1.1160)  & & & & \\[\dadd]
$(-6,  -4)$ & (0.0, 1.1030)  & & & & \\[\dsub]
          & (0.0, 1.1028)  & & & & \\[\dadd]
\hline
$(-10, -1)$ & (0.0, 1.3200)     & (0.1202, 1.3200)  & (0.2844, 1.3200)  & & \\[\dsub]
          & (0.0, 1.3199)     & (0.1167, 1.3199)  & (0.2743, 1.3200)  & & \\[\dadd]
$(-9,  -2)$ & (0.0, 1.2851)     & (0.1773, 1.2851)  & & & \\[\dsub]
          & (0.0, 1.2851)     & (0.1670, 1.2873)  & & & \\[\dadd]
$(-8,  -3)$ & (0.0, 1.2590)     & (0.0895, 1.2590)  & & & \\[\dsub]
          & (0.0, 1.2590)     & (0.0817, 1.2609)  & & & \\[\dadd]
$(-7,  -4)$ & (0.0, 1.2416)  & & & & \\[\dsub]
          & (0.0, 1.2414)  & & & & \\[\dadd]
$(-6,  -5)$ & (0.0, 1.2329)  & & & & \\[\dsub]
          & (0.0, 1.2328)  & & & & \\[\dadd]
\hline
$(-10, -2)$ & (0.0776, 1.4471)  & (0.2370, 1.4471)  & & & \\[\dsub]
          & (0.0748, 1.4472)  & (0.2271, 1.4485)  & & & \\[\dadd]
$(-9,  -3)$ & (0.0, 1.4138)     & (0.1356, 1.4138)  & & & \\[\dsub]
          & (0.0, 1.4138)     & (0.1266, 1.4161)  & & & \\[\dadd]
$(-8,  -4)$ & (0.0551, 1.3901)  & & & & \\[\dsub]
          & (0.0492, 1.3913)  & & & & \\[\dadd]
$(-7,  -5)$ & (0.0, 1.3758)  & & & & \\[\dsub]
          & (0.0, 1.3758)  & & & & \\[\dadd]
\hline
$(-10, -3)$ & (0.0, 1.5806)     & (0.1909, 1.5806)  & & & \\[\dsub]
          & (0.0, 1.5806)     & (0.1818, 1.5826)  & & & \\[\dadd]
$(-9,  -4)$ & (0.0, 1.5497)     & (0.0964, 1.5497)  & & & \\[\dsub]
          & (0.0, 1.5497)     & (0.0895, 1.5516)  & & & \\[\dadd]
$(-8,  -5)$ & (0.0, 1.5291)  & & & & \\[\dsub]
          & (0.0, 1.5291)  & & & & \\[\dadd]
$(-7,  -6)$ & (0.0, 1.5188)  & & & & \\[\dsub]
          & (0.0, 1.5188)  & & & & \\[\dadd]
\hline
$(-10, -4)$ & (0.0, 1.7216)     & (0.1463, 1.7216)  & & & \\[\dsub]
          & (0.0, 1.7216)     & (0.1387, 1.7236)  & & & \\[\dadd]
$(-9,  -5)$ & (0.0595, 1.6938)  & & & & \\[\dsub]
          & (0.0544, 1.6950)  & & & & \\[\dadd]
$(-8,  -6)$ & (0.0, 1.6772)  & & & & \\[\dsub]
          & (0.0, 1.6772)  & & & & \\[\dadd]
\hline
$(-10, -5)$ & (0.0, 1.8713)     & (0.1043, 1.8713)  & & & \\[\dsub]
          & (0.0, 1.8713)     & (0.0985, 1.8729)  & & & \\[\dadd]
$(-9,  -6)$ & (0.0, 1.8475)  & & & & \\[\dsub]
          & (0.0, 1.8475)  & & & & \\[\dadd]
$(-8,  -7)$ & (0.0, 1.8356)  & & & & \\[\dsub]
          & (0.0, 1.8357)  & & & & \\[\dadd]
\hline
$(-10, -6)$ & (0.0644, 2.0308)  & & & & \\[\dsub]
          & (0.0604, 2.0318)  & & & & \\[\dadd]
$(-9,  -7)$ & (0.0, 2.0118)  & & & & \\[\dsub]
          & (0.0, 2.0119)  & & & & \\[\dadd]
\hline
$(-10, -7)$ & (0.0, 2.2015)  & & & & \\[\dsub]
          & (0.0, 2.2016)  & & & & \\[\dadd]
$(-9,  -8)$ & (0.0, 2.1881)  & & & & \\[\dsub]
          & (0.0, 2.1881)  & & & & \\[\dadd]
\hline
$(-10, -8)$ & (0.0, 2.3845)  & & & & \\[\dsub]
          & (0.0, 2.3846)  & & & & \\[\dadd]
\hline
$(-10, -9)$ & (0.0, 2.5809)  & & & & \\[\dsub]
          & (0.0, 2.5810)  & & & & \\[\dadd]
\end{tabular}
\end{center}
\label{alldps}
\end{table}

\begin{thebibliography}{10}
\bibitem{fnscs} The spin coherent state $\ket{\nhat}$, where $\nhat$ is the
direction with spherical polar coordinates $(\tta,\phi)$, is the state of
maximum spin projection along $\nhat$: $\bS\cdot\nhat\ket{\nhat} = S \ket{\nhat}$.
\bibitem{note4}
Consider the polynomial $(x - a_1)(x - a_2 )\cdot \cdot \cdot (x -
a_n)$, where the $a_i$ are all positive. The coeffcient of
$x^{n-k}$ is $(-1)^k$ times the symmetric sum $\sum
a_{i_1}a_{i_2}\cdot\cdot\cdot a_{i_k}$ of all distinct $k$-fold
products of the $a$'s.
\bibitem{wwpc} The considerations of this subsection were prompted by
a private communication from Dr. Wernsdorfer, informing us that 
numerical diagonalization with the \Mn12 parameters revealed that in
addition to being quenched on the $h_z$ axis, the splitting of the
$m=-10$ and $m'=9$ states was also quenched at a very close by point.
\bibitem{fnabc} Hence, $a = 1.067 \times 10^{-2}$. It seems safe to
assume that $c \ll 1$, so that $b \simeq 3\times 10^{-4}$.
\end{thebibliography}
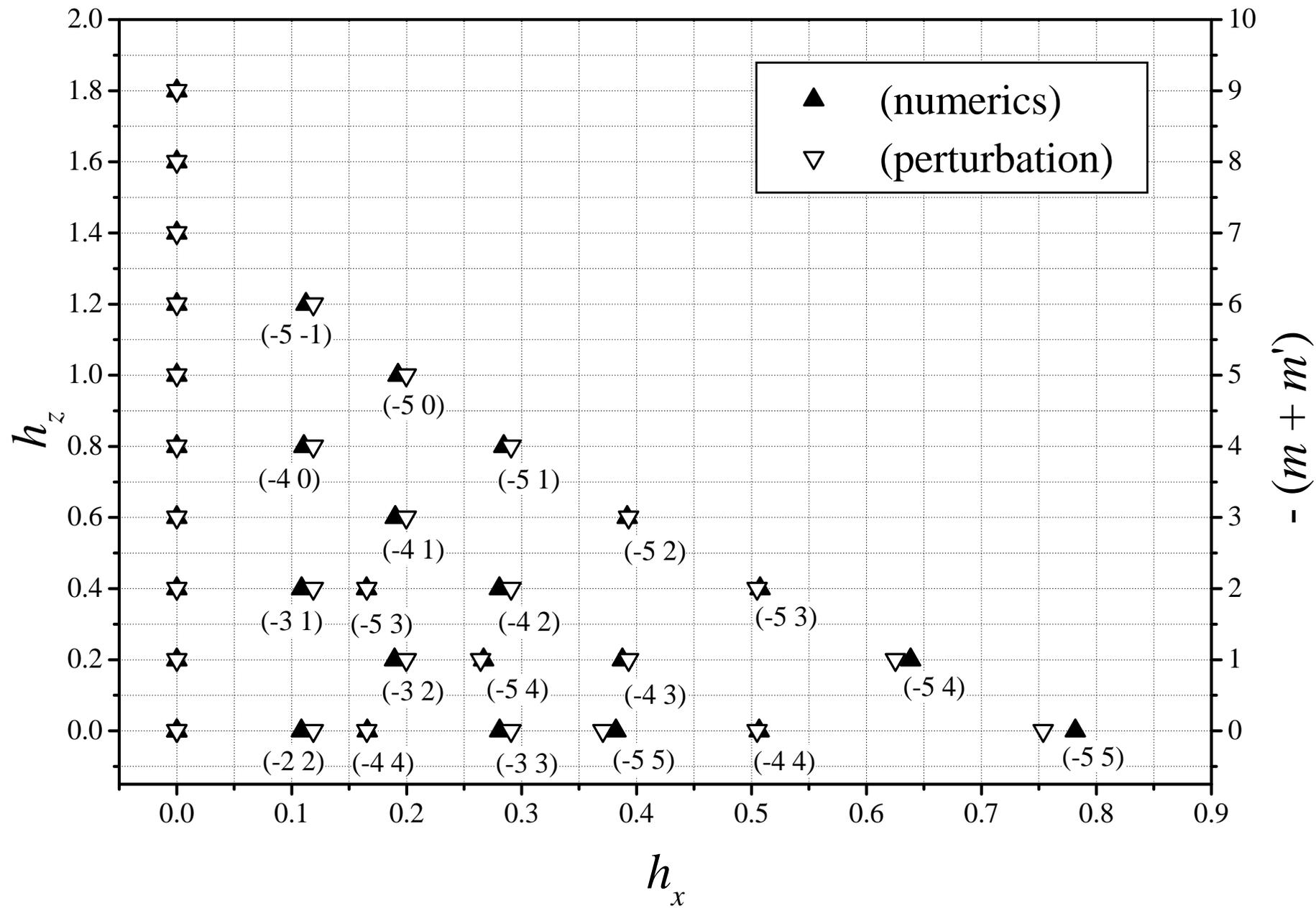
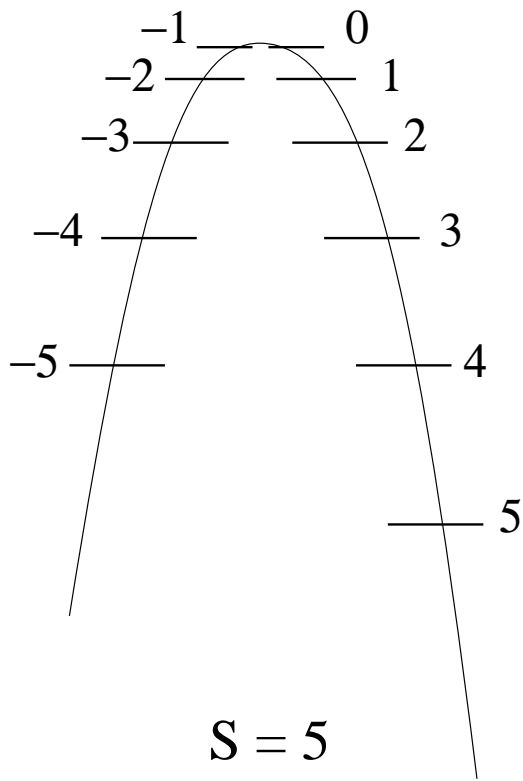
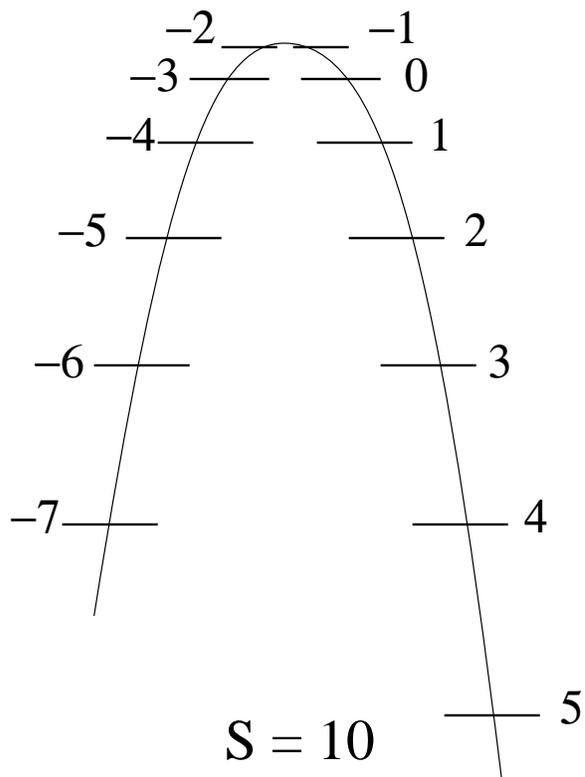
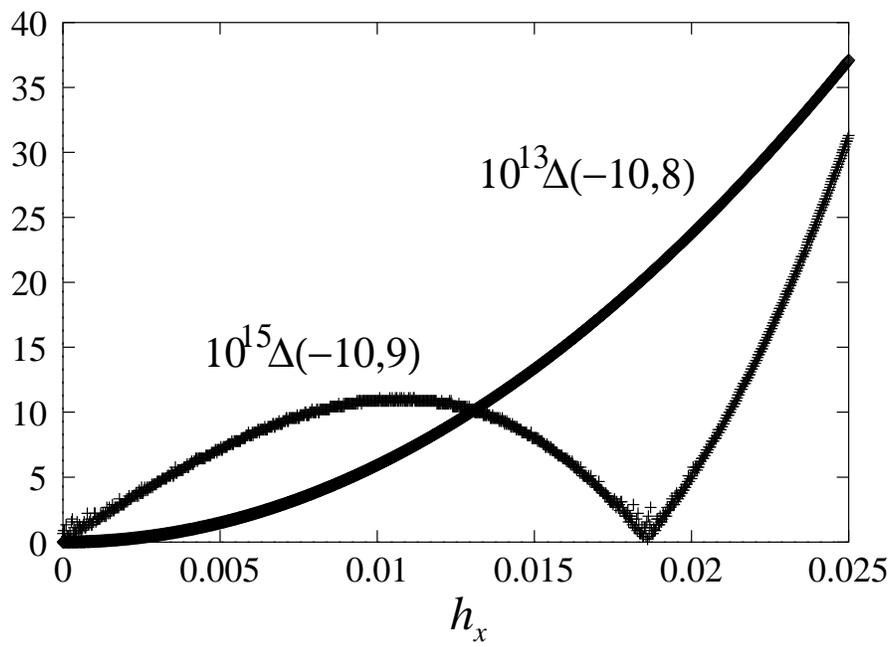
\end{document}